\documentclass[twocolumn]{pasj00}

\def\gsim{\mathrel{\rlap{\lower 4pt \hbox{\hskip 1pt $\sim$}}\raise 1pt
\hbox {$>$}}}
\def\lsim{\mathrel{\rlap{\lower 4pt \hbox{\hskip 1pt $\sim$}}\raise 1pt
\hbox {$<$}}}

\begin{document}
\SetRunningHead{M. Yamanaka et al.}{U Sco 2010}
\Received{}
\Accepted{}




\title{Early Spectroscopy of the 2010 Outburst of U Scorpii}

\author{Masayuki \textsc{Yamanaka}\altaffilmark{1,2},
	Makoto \textsc{Uemura}\altaffilmark{2}, 
        Koji S. \textsc{Kawabata}\altaffilmark{2},
	Mitsugu  \textsc{Fujii}\altaffilmark{3}, \\
	Kenji \textsc{Tanabe}\altaffilmark{4},
	Kazuyoshi  \textsc{Imamura}\altaffilmark{4},
	Tomoyuki \textsc{Komatsu}\altaffilmark{1}, 
	Akira \textsc{Arai}\altaffilmark{5}, 
	Mahito \textsc{Sasada}\altaffilmark{1}, 
	Ryosuke \textsc{Itoh}\altaffilmark{1}, \\ 
	Tatsuya \textsc{Harao}\altaffilmark{1},  
 	Nanae \textsc{Kunitomi}\altaffilmark{4},	
	Osamu \textsc{Nagae}\altaffilmark{1}, 
	Mikiha \textsc{Nose}\altaffilmark{4}, 
	Takashi \textsc{Ohsugi}\altaffilmark{2}, \\
	Takako \textsc{Okushima}\altaffilmark{1}, 
	Kiyoshi \textsc{Sakimoto}\altaffilmark{1}, 
	Michitoshi \textsc{Yoshida}\altaffilmark{2}
}

\altaffiltext{1}{Department of Physical Science, Hiroshima University, Kagamiyama 1-3-1, Higashi-Hiroshima 739-8526, Japan; myamanaka@hiroshima-u.ac.jp} 
\altaffiltext{2}{Hiroshima Astrophysical Science Center, Hiroshima University, Higashi-Hiroshima, Hiroshima 739-8526, Japan} 
\altaffiltext{3}{Fujii Kurosaki Observatory, Kurashiki, Okayama, Japan} 
\altaffiltext{4}{Department of Biosphere-Geosphere System Science, Faculty of
Informatics, Okayama University of Science, 1-1 Ridai-cho, Okayama, 
Okayama 700-0005, Japan} 
\altaffiltext{5}{Koyama Astronomical Observatory, Kyoto Sangyo University, Motoyama, Kamigamo, Kita-ku, Kyoto 603-8555, Japan} 



\KeyWords{Spectroscopy --- Star: novae --- Novae: individual: U Sco} 

\maketitle

\begin{abstract}

 We present early spectroscopy of the recurrent nova U~Sco during the
 outburst in 2010. We successfully obtained time-series spectra at
 $t_{\rm d}=$0.37--0.44~d, where $t_{\rm d}$ denotes the time
 from the discovery of the present outburst. This is the first 
 time-resolved spectroscopy on the first night of U Sco outbursts.
 At $t_{\rm d}\sim 0.4$~d the H$\alpha$ line consists
 of a blue-shifted ($-5000$ km s$^{-1}$) narrow absorption component
 and a wide emission component having triple peaks, a blue ($\sim -3000$ 
 km s$^{-1}$), a central ($\sim 0$ km s$^{-1}$) and a red
 ($\sim +3000$ km s$^{-1}$) ones. The blue and red peaks developed
 more rapidly than the central one during the first night. 
 This rapid variation would be caused by the growth of aspherical 
 wind produced during the earliest stage of the outburst. 
 At $t_{\rm d}=1.4$~d the H$\alpha$ line has a nearly flat-topped
 profile with weak blue and red peaks at $\sim \pm 3000$ km s$^{-1}$.
 This profile can be attributed to a nearly spherical shell, while
 the asphericity growing on the first night still remains.
 The wind asphericity is less significant after $t_{\rm d}=9$ d.
\end{abstract}


\section{Introduction}
 U Sco is a recurrent nova (RN) whose outburst occurs every 8--12 years. 
 The recurrence time of U Sco is the shortest among all known RNe.
 U Sco is also known as a very fast nova defined 
 by $t_2<10$~d, where $t_2$ is an elapsed time when the object faded by 2~mag 
 from the outburst maximum \citep{Payne-Gaposchkin1958}. 
 Because of the rapid evolution, the earliest phase ($\lesssim 1$~d)
 of the outburst of U Sco is still poorly known in spite of its 
 relatively short recurrence period.

 \citet{Schaefer1995} performed photometry of U~Sco in its quiescent
 phase and found that it is an eclipsing binary having an orbital
 period of 1.23~d. The eclipse was also observed in the outburst in 
 1999 \citep{Matusmoto2003}. \citet{Hachisu2000} successfully reproduced 
 the light curve including the eclipse,
 using their theoretical model with a very massive white dwarf
 (WD) close to the Chandrasekhar limit.
 Hence, U~Sco is suggested to be a system in the final stage of
 binaries evolving to Type Ia supernovae \citep{Hachisu1999, Hachisu2008}.

 Thanks to the network for circulating the discovery reports and observations of
 transient objects (see \cite{Kato2004} for review), observations just
 after the outburst maximum of U~Sco were first performed in 1999 
 \citep{Munari1999,Anupama2000,Iijima2002}.
 \citet{Munari1999} and \citet{Iijima2002} reported that the
 Balmer emission lines had triple peaks where the red peak
 is stronger than the blue and central ones at $t_{\rm p}\simeq +0.65$~d, 
 where $t_{\rm p}$ is the elapsed time from the peak of the outburst.
 The width of the emission lines gradually decreased with time.
 \citet{Anupama2000} took an earlier spectrum at $t_{\rm p}=+0.45$~d,
 in which the red peak of H$\alpha$ was weaker than the central one.
 These observations imply that a rapid change in the line profile
 would have occurred only within a few hours after the maximum brightness.
 This could be attributed to a temporal change in the structure of
 the earliest wind.
 A time-series spectroscopy in the earliest phase is required
 to probe the nature of the U Sco outburst.

 
 We successfully obtained spectroscopic and photometric data of U~Sco in 
 the earliest phase just after the outburst discovery in 2010. In this 
 Letter, we present the results of the spectroscopic observations
 in an early stage of the outburst. The results of photometric and
 late-time spectroscopic observations will be published in a forthcoming paper.

\section{Observations and Reduction}


\begin{table}
\caption{Log of Spectroscopic Observations}
\begin{center}
\begin{tabular}{cccc}
\hline
Date (UT)&$t_{\rm d}$$^{a}$&Inst.$^{b}$&Phase$^{c}$  \\
\hline \hline
Jan.~28.84&+0.37&1&0.05  \\
Jan.~28.84&+0.37&2&0.05  \\
Jan.~28.86&+0.39&3&0.07  \\
Jan.~28.86&+0.39&1&0.07  \\
Jan.~29.85&+1.38&1&0.87  \\
Jan.~29.85&+1.38&2&0.87 \\
Feb.~2.79&+5.32&3&0.08  \\
Feb.~2.86&+5.38&1&0.13  \\
Feb.~3.86&+6.39&3&0.95  \\
Feb.~3.87&+6.40&1&0.95  \\
Feb.~4.85&+7.38&3&0.75  \\
Feb.~4.87&+7.40&1&0.77  \\
Feb.~5.83&+8.36&3&0.55  \\
Feb.~5.83&+8.36&1&0.77  \\
Feb.~6.85&+9.38&3&0.38  \\
Feb.~13.83&+16.36&3&0.05  \\
Feb.~19.86&+22.39&3&0.95  \\
\hline
\end{tabular}
\end{center}
{\bf Note.} $^{a}$The time from the discovery of the current outburst (2010 Jan 28.4385 UT).
$^{b}$The used instrument (1=FBSPEC~{\sc II}, 2=OUS/SBIG-DSS7,3=Kanata/HOWPol).
$^{c}$The orbital phase \citep{Schaefer1995}.
\end{table}

 \begin{figure*}
 \begin{center}
 \begin{tabular}{c}
 \includegraphics[scale=0.7]{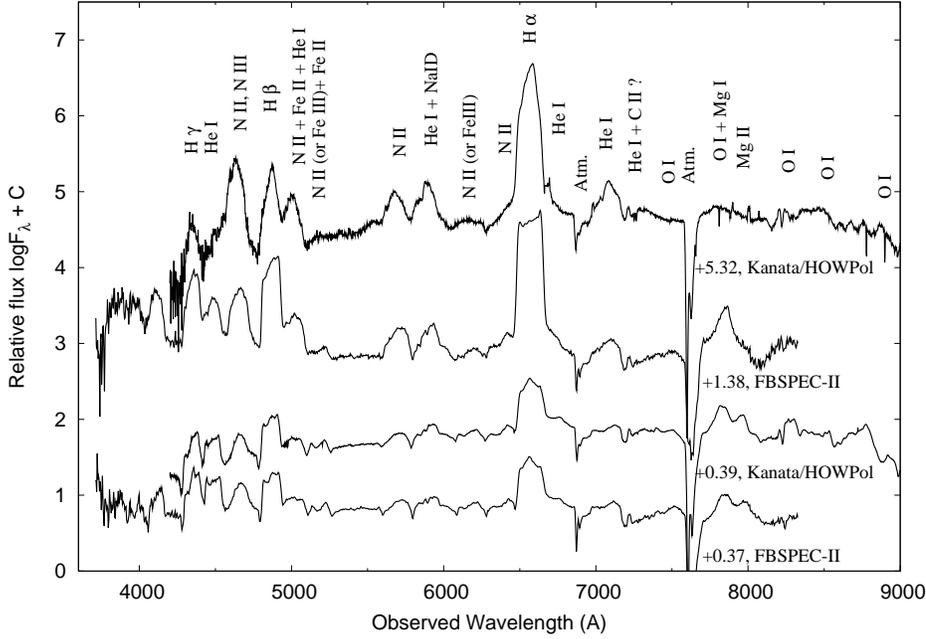}
 \end{tabular}
 \caption{Early phase spectra of U Sco . These spectra were obtained in series,
 from bottom to top, on Jan. 28.84, 28.86, 29.85 and Feb. 2.79.
 The line identification is based on \citet{Munari1999,Anupama2000,Iijima2002}.}
 \end{center}
 \end{figure*}

 The tenth outburst of U~Sco was independently discovered at $V=8.05$
 on 2010 Jan 28.4385 UT by S. Dvorak and at visual magnitude of 8.8 by
 B. G. Harris \citep{Schaefer2010a}.  In this Letter, we use the epoch
 , $t_{\rm d}$, which is the time since the discovery date by S. Dvorak.
 \footnote{\citet{Schaefer2010b} suggest that the peak of the 2010 
 outburst would be Jan 28.19 with an uncertainty of 0.07 day from 
 the similarity of the light curves in every eruption of RNe.}.  
 We started spectroscopic observations of the outburst just 
 after the discovery.

 We performed spectroscopic observations of U~Sco on 8 nights from
 2010 Jan 28.86 to Feb 19.86 using HOWPol (Hiroshima One-shot 
 Wide-field Polarimeter ; \cite{Kawabata2008}) installed to the 
 1.5-m Kanata telescope at Higashi-Hiroshima Astronomical Observatory (HHAO). 
 The wavelength coverage was 4200--9000 \AA\ and the spectral resolution
 was $R=\lambda /\Delta\lambda \sim 400$ at 6000 \AA.
 On the first night, we performed time-resolved spectroscopy during  
 Jan 28.84 -- 28.91 with HOWPol.
 We also performed low-resolution spectroscopy with FBSPEC~{\sc II} 
 attached to a 40 cm reflector from Jan 
 28.84 to Feb 5.83 at Fujii Kurosaki Observatory (FKO).
 The wavelength coverage was 3800-8400 \AA\ and the spectral resolution
 was $R=\lambda /\Delta\lambda \sim 500$ at 6000 \AA.
 Additional spectra were obtained with SBIG DSS7 attached to a 28 cm 
 reflector on 2010 Jan 28.840 and 29.846 at the observatory in Okayama 
 University of Science (OUS). The wavelength coverage was 4200--8200 \AA\
 and the spectral resolution was $R=\lambda /\Delta\lambda \sim 400$ at 
 6000 \AA. The log of our spectroscopic observations is summarized in 
 table 1. The data reductions of these data were performed according 
 to a standard procedure using IRAF. The wavelength calibration was 
 performed using the sky emission lines taken in the same frame.

 \section{Results}

 \subsection{Overall Properties of Early Spectra}

  In figure 1, we show the spectra obtained on Jan 28.84 
 ($t_{\rm d}=0.37$~d), 28.86 ($t_{\rm d}=0.39$~d), 29.85 
 ($t_{\rm d}=1.38$~d) and Feb 2.79 ($t_{\rm d}=5.32$~d). 
 We performed line identification by comparing with previous studies
 \citep{Munari1999, Anupama2000, Iijima2002} and show them in figure 1.
 They are characterized by some strong emission lines having large FWZI 
 (Full-Width at the Zero-Intensity) of $\sim 11000$ km~s$^{-1}$. 

 Blue-shifted narrow absorption lines are accompanied with
 some strong emission lines at $t_{\rm d}=0.37$ and $0.39$~d.
 They are considerably weakened by $t_{\rm d}=1.38$~d and
 diminished by $t_{\rm d}=5.32$~d.
 The velocity of the absorption components are about $-5000$,
 $-4600$ and $-4500$ km s$^{-1}$ for Balmer series,
 He~{\sc i} and N~{\sc ii} lines, respectively. 

 The profile of the H$\alpha$ emission line consists of a central,
 red, and blue components at $t_{\rm d}\simeq 0.4$~d.  The red and
 blue components have peaks at $\sim \pm 3000$ km s$^{-1}$, as shown
 in figure 2.
 The red component is stronger than the blue one.
 A similar profile is seen in the spectrum in the 1999 outburst at
 $t_{\rm p}=0.45$~d, which is the earliest spectrum taken in that
 outburst \citep{Anupama2000}.
 In figure 3, we show the nightly variation of the whole H$\alpha$
 line in a logarithmic scale.  At $t_{\rm d}=1.38$~d, the line profile
 is described as a nearly flat-topped one with weak blue and red peaks
 at $\pm 3000$ km s$^{-1}$.  The red peak is still higher than the blue
 one, as observed at $t_{\rm d}\sim 0.4$~d.  Then, the peak wavelength
 gradually shifts to the rest one between $t_{\rm d}\sim 5$ and 9~d.
 At $t_{\rm d}=9.38$~d, the H$\alpha$ line had a rather symmetric
 profile consisting of the flat-topped component and a narrower peak
 one.  In the spectra at $t_{\rm d}=16.36$ and $22.39$~d, the flat-topped
 component diminishes and the line profile is characterized with a
 single peak having a possible spread wing.

 The H$\beta$, He~{\sc i}$\lambda$5876 and He~{\sc i}$\lambda$7065
 emission lines exhibit the same behavior as H$\alpha$, except for
 the relative intensity of the blue and red peaks.  The ratio of
 their blue and red peaks relative to the central one is higher
 than that of H$\alpha$, as can be seen in figure~2.


 \begin{figure}
 \begin{center}
 \begin{tabular}{c}
 \includegraphics[scale=0.7]{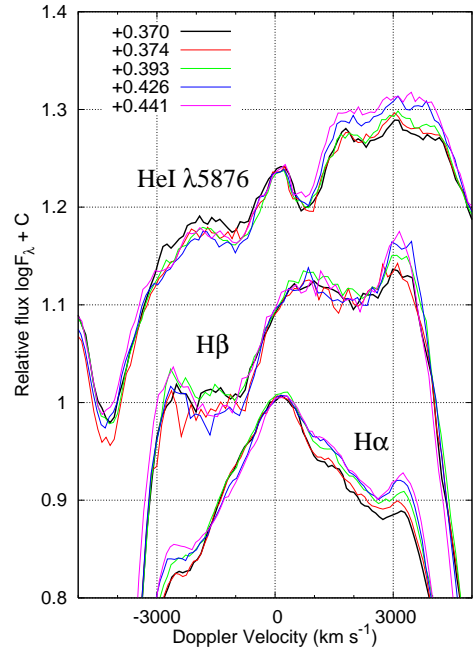}
 \end{tabular}
 \caption{The intra-night variability of the peak profiles of H$\alpha$, H$\beta$
 and He~{\sc i}$\lambda$5876 in $t_{\rm d}=0.37$--$0.44$~d. To emphasize the 
 intra-night increasing of the blue and red components, the spectra 
 are normalized at the rest wavelength of each line. 
 The total intensity of each emission line gradually becomes
 stronger by $\sim 1$ \% within two hours from $t_{\rm d}=0.37$ 
 to $0.44$~d.
 } 
 \end{center} 
 \end{figure}

 \begin{figure}
 \begin{center}
 \begin{tabular}{c}
 \includegraphics[scale=0.85]{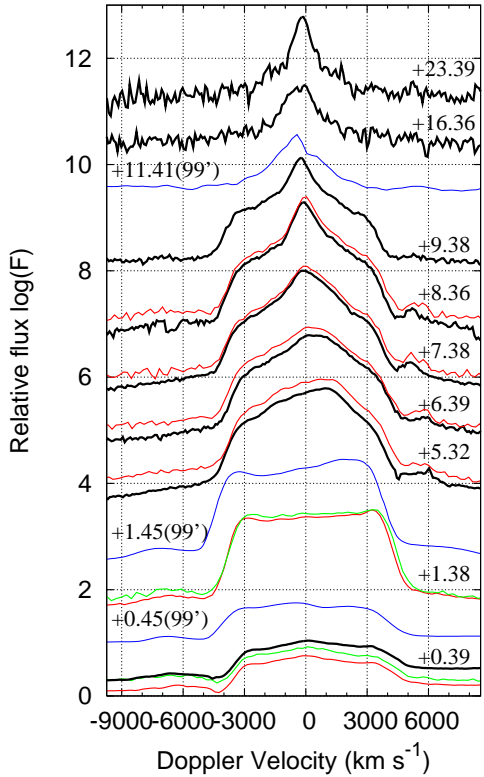}
 \end{tabular}
 \caption{The inter-night variability in the line profile of the H$\alpha$. 
  The black, red and green lines show the spectra obtained at HHAO, FKO and 
  OUS, respectively. The blue lines show the spectra from \citet{Anupama2000} 
  in order to compare the profile for 2010 outburst. The values written
  at the left and right sides of the figure mean the days from the outburst 
  maximum in 1999 and the days from the discovery date in 2010, respectively.
}
 \end{center} 
 \end{figure}

 \subsection{Spectral Variation on the First Night}

 As shown in figure~2, the intensity of the blue and red components of
 H$\alpha$ increases more rapidly than that of the central one on the 
 first night. The growth of the red component is apparently smooth, 
 while that of the blue component may be more complicated, or weaker 
 than the red one.
 Similarly, the H$\beta$ and He~{\sc i}$\lambda$5876
 lines show a rapid increase in their red components.
 These rapidly-evolving components are seen not only around
 the red peak, but also in a rather wide range of velocity
 from 1000--4000~km~s$^{-1}$ in H$\alpha$ and He~{\sc i}$\lambda$5876.

 In the H$\alpha$ line, the central peak is stronger than the red and
 blue ones on the first night, while the whole profile becomes a 
 flat-topped one by $t_{\rm d}=1.38$~d. Hence, it is expected that the
 increase of the blue and red components
 continue until they became comparable to that of the central component.
 Thus, our observation on the first night traces the growing phase
 of the blue and red peaks. The observation epoch would correspond
 to the period of time that was overlooked in the case of the 1999
 outburst, namely, between $t_{\rm p}=0.45$~d \citep{Anupama2000}
 and $0.65$~d \citep{Iijima2002} in the 1999 outburst.

\section{Discussion}
 
 The intra-night variation of the blue and red components shown
 in figure~2 suggests that the emitting region has an aspherical 
 shape that grows rapidly on the first night. \citet{Gill1999}
 calculated line profiles using the models with aspherical
 structure of nova ejecta, like equatorial and tropical rings or
 polar caps, which generate blue and red peaks in the line profile.
 We propose that aspherical winds, such as
 suggested in \citet{Gill1999}, is the origin of the red
 (and blue) components. A bipolar jet-like structure has been
 suggested for U~Sco to explain its large expansion velocity (see
 \cite{MKato2003} and reference therein). A well-collimated narrow
 jet-like structure is expected to generate emission lines with a narrow
 velocity range. Since the rapidly-growing red components extend
 in a wide velocity range of 1000--4000~km~s$^{-1}$ (see \S 3.2), the 
 observed line variation favors a wind-like structure with a large 
 opening angle, rather than a well-collimated jet-like structure 
 with a small opening angle. In contrast to the blue and red 
 components, the central component, slowly growing relative to the
 red one, is probably attributed to a slower, nearly spherical wind.

 The fact that the red component is stronger
 than the blue one on the first night implies that the 
 wind is not only aspherical, but also non-axisymmetric. On the 
 other hand, the blue component could also be supressed due to 
 superposition of blue-shifted absorption components.
 The latter is plausible because blue-shifted absorption 
 components are clearly seen at that phase. In addition, 
 it is interesting that the orbital phase is close to $zero$ 
 during our observation; Phase$=0.05$ at $t_{\rm d}=0.37$, as 
 shown in table~1. This suggests that the secondary star is at 
 near side and might give a partial occultation or a perturbation 
 of the coming wind, leading to weaker blue components. This 
 scenario is plausible only when the secondary star affects 
 the velocity field of the ejecta significantly. 

 Another noteworthy feature of the line profile on the first night
 is that H$\beta$ and He~{\sc i}$\lambda$5876 have strong red
 components relative to H$\alpha$. In other words, the central
 component of H$\beta$ and He~{\sc i}$\lambda$5876 is relatively
 weak. This would reflect the difference in temperature of the 
 emitting region. Typically, H$\beta$ and 
 He~{\sc i}$\lambda$5876 may become stronger relatively to 
 H$\alpha$ in the case of higher electron temperature.
 The temperatures of the excitations of H$\beta$ and 
 He~{\sc i}$\lambda$5876 are higher than that of H$\alpha$. 
 The observed line profiles suggest that the possible
 asymmetric winds causing the red and blue components
 are highly excited compared with the  slower, nearly
 spherical wind causing the central one.


 The asymmetry of the nova ejecta has been discussed in many
 literature (e.g., \cite{MacDonald1986, Livio1990, Porter1998,
 Scott2000, Gill2000, Harman2003, Kawabata2006}).
 In the 1999 outburst in U Sco, a significant polarization change
 across the H$\alpha$ emission line was found at $t_{\rm p}=0.3$~d
 and then diminished by $t_{\rm p}=2.2$~d \citep{Ikeda2000}.
 This result suggests that the earliest wind was asymmetric,
 which is consistent with the present work.

 At $t_{\rm d}=1.38$~d, the line profile of H$\alpha$ has a nearly
 flat-topped structure with weak blue and red peaks. This profile
 implies that the H$\alpha$ emission is from a nearly spherical
 shell on the second night, while the asphericity growing on the
 first night still remains.
 At $t_{\rm d}\sim 9$~d, the profile of the H$\alpha$ line consists
 of the flat-topped component and another Gaussian component centered
 at the rest wavelength. The asymmetry in the line profile gradually 
 becomes insignificant by $t_{\rm d}\sim 9$~d. This indicates the 
 asphericity of the wind is less significant by this time.
 Since the deceleration of the flat-topped component is not remarkable
 between $t_{\rm d}=0.39$ and 9.38~d, the spherical shell is likely to 
 expand freely from the first night. The Gaussian component would be 
 attributed to the decelerating nova wind near the white dwarf.
 By $t_{\rm d }= 16.36$~d, the flat-topped component diminishes.
 It is noted that the single-peaked H$\alpha$ line after $t_{\rm d}=16$~d
 is similar to that in the same epoch of the 1987 outburts 
 \citep{Sekiguchi1988}, while not consistent with those in 
 1979 and 1999 outbursts \citep{Barlow1981, Munari1999}. 
 This may suggest that the geometry (shape and/or its axis) of the
 outburst is not always same among two consecutive outbursts. 
 Additionally, the observed maximum of the outbursts in 1999 and 2010
 are estimated to be the phase $\sim0.41$ and $0.05$, respectively. 
 Assuming that the geometry of the outburst is not axi-symmetric and 
 its shape depends on the binary phase at the onset of the outburst, 
 we expect the apparent nature could change along different lines of sight. 
 If this is the case, the high-velocity winds could be seen at the 
 outburst begun around the phase of $\sim 0.4$ \citep{Munari1999}.

 \section{Conclusion}

 We obtained the time-resolved spectra on the first night after the
 discovery of the 2010 outburst of U Sco, and found the rapid evolution 
 of the profile of the H$\alpha$ emission line. This is the first 
 observational sample revealing the intra-night variability of the spectral
 lines just 
 after the maximum brightness. The H$\alpha$ emission line shows 
 three components, which are, center, blue, and red ones on the first night.
 The blue and red components rapidly evolved, which would be caused
 by aspherically-structured winds. The line profile becomes rather
 symmetric and we cannot find any clear evidence for the wind asphericity
 after $t_{\rm d}=9$~d.

 \vspace{20pt}

 We would like to thank G. C. Anupama for permission to
 use their spectra of the 1999 outburst.
 This research has been supported in part by the Grant-in-Aid for 
 Scientific Research from JSPS (17684004, 20740107, 21018007) .
 M.Y. and M.S. have been supported by the JSPS Research Fellowship
 for Young Scientists.


\begin{thebibliography}{25}
\expandafter\ifx\csname natexlab\endcsname\relax\def\natexlab#1{#1}\fi

\bibitem[{{Anupama} \& {Dewangan}(2000)}]{Anupama2000}
{Anupama}, G.~C., \& {Dewangan}, G.~C. 2000, \aj, 119, 1359

\bibitem[{{Barlow} {et~al.}(1981){Barlow}, {Brodie}, {Brunt}, {Hanes}, {Hill},
  {Mayo}, {Pringle}, {Ward}, {Watson}, {Whelan}, \& {Willis}}]{Barlow1981}
{Barlow}, M.~J., {et~al.} 1981, \mnras, 195, 61

\bibitem[{{Gill} \& {O'Brien}(1999)}]{Gill1999}
{Gill}, C.~D., \& {O'Brien}, T.~J. 1999, \mnras, 307, 677

\bibitem[{{Gill} \& {O'Brien}(2000)}]{Gill2000}
---. 2000, \mnras, 314, 175

\bibitem[{{Hachisu} {et~al.}(2000){Hachisu}, {Kato}, {Kato}, \&
  {Matsumoto}}]{Hachisu2000}
{Hachisu}, I., {Kato}, M., {Kato}, T., \& {Matsumoto}, K. 2000, \apjl, 528, L97

\bibitem[{{Hachisu} {et~al.}(2008){Hachisu}, {Kato}, \& {Nomoto}}]{Hachisu2008}
{Hachisu}, I., {Kato}, M., \& {Nomoto}, K. 2008, \apj, 679, 1390

\bibitem[{{Hachisu} {et~al.}(1999){Hachisu}, {Kato}, {Nomoto}, \&
  {Umeda}}]{Hachisu1999}
{Hachisu}, I., {Kato}, M., {Nomoto}, K., \& {Umeda}, H. 1999, \apj, 519, 314

\bibitem[{{Harman} \& {O'Brien}(2003)}]{Harman2003}
{Harman}, D.~J., \& {O'Brien}, T.~J. 2003, \mnras, 344, 1219

\bibitem[{{Iijima}(2002)}]{Iijima2002}
{Iijima}, T. 2002, \aap, 387, 1013

\bibitem[{{Ikeda} {et~al.}(2000){Ikeda}, {Kawabata}, \& {Akitaya}}]{Ikeda2000}
{Ikeda}, Y., {Kawabata}, K.~S., \& {Akitaya}, H. 2000, \aap, 355, 256

\bibitem[{{Kato} \& {Hachisu}(2003)}]{MKato2003}
{Kato}, M., \& {Hachisu}, I. 2003, \apjl, 587, L39

\bibitem[{{Kato} {et~al.}(2004){Kato}, {Uemura}, {Ishioka}, {Nogami},
  {Kunjaya}, {Baba}, \& {Yamaoka}}]{Kato2004}
{Kato}, T., {Uemura}, M., {Ishioka}, R., {Nogami}, D., {Kunjaya}, C., {Baba},
  H., \& {Yamaoka}, H. 2004, \pasj, 56, 1

\bibitem[{{Kawabata} {et~al.}(2006){Kawabata}, {Ohyama}, {Ebizuka}, {Takata},
  {Yoshida}, {Isogai}, {Norimoto}, {Okazaki}, \& {Saitou}}]{Kawabata2006}
{Kawabata}, K.~S., {et~al.} 2006, \aj, 132, 433

\bibitem[{{Kawabata} {et~al.}(2008){Kawabata}, {Nagae}, {Chiyonobu}, {Tanaka},
  {Nakaya}, {Suzuki}, {Kamata}, {Miyazaki}, {Hiragi}, {Miyamoto}, {Yamanaka},
  {Arai}, {Yamashita}, {Uemura}, {Ohsugi}, {Isogai}, {Ishitobi}, \&
  {Sato}}]{Kawabata2008}
{Kawabata}, K.~S., {et~al.} 2008, in Society of Photo-Optical Instrumentation
  Engineers (SPIE) Conference Series, Vol. 7014, Society of Photo-Optical
  Instrumentation Engineers (SPIE) Conference Series

\bibitem[{{Livio} {et~al.}(1990){Livio}, {Shankar}, {Burkert}, \&
  {Truran}}]{Livio1990}
{Livio}, M., {Shankar}, A., {Burkert}, A., \& {Truran}, J.~W. 1990, \apj, 356,
  250

\bibitem[{{MacDonald}(1986)}]{MacDonald1986}
{MacDonald}, J. 1986, \apj, 305, 251

\bibitem[{{Matsumoto} {et~al.}(2003){Matsumoto}, {Kato}, \&
  {Hachisu}}]{Matusmoto2003}
{Matsumoto}, K., {Kato}, T., \& {Hachisu}, I. 2003, \pasj, 55, 297

\bibitem[{{Munari} {et~al.}(1999){Munari}, {Zwitter}, {Tomov}, {Bonifacio},
  {Molaro}, {Selvelli}, {Tomasella}, {Niedzielski}, \& {Pearce}}]{Munari1999}
{Munari}, U., {et~al.} 1999, \aap, 347, L39

\bibitem[{{Payne-Gaposchkin}(1958)}]{Payne-Gaposchkin1958}
{Payne-Gaposchkin}, C. 1958, \jrasc, 52, 92

\bibitem[{{Porter} {et~al.}(1998){Porter}, {O'Brien}, \& {Bode}}]{Porter1998}
{Porter}, J.~M., {O'Brien}, T.~J., \& {Bode}, M.~F. 1998, \mnras, 296, 943

\bibitem[{{Schaefer} {et~al.}(2010{\natexlab{a}}){Schaefer}, {Harris},
  {Dvorak}, {Templeton}, \& {Linnolt}}]{Schaefer2010a}
{Schaefer}, B.~E., {Harris}, B.~G., {Dvorak}, S., {Templeton}, M., \&
  {Linnolt}, M. 2010{\natexlab{a}}, \iaucirc, 9111, 1

\bibitem[{{Schaefer} \& {Ringwald}(1995)}]{Schaefer1995}
{Schaefer}, B.~E., \& {Ringwald}, F.~A. 1995, \apjl, 447, L45+

\bibitem[{{Schaefer} {et~al.}(2010{\natexlab{b}}){Schaefer}, {Pagnotta},
  {Xiao}, {Darnley}, {Bode}, {Harris}, {Dvorak}, {Menke}, {Linnolt},
  {Templeton}, {Henden}, {Pojma{\'n}ski}, {Pilecki}, {Szczygiel}, \&
  {Watanabe}}]{Schaefer2010b}
{Schaefer}, B.~E., {et~al.} 2010{\natexlab{b}}, ArXiv e-prints

\bibitem[{{Scott}(2000)}]{Scott2000}
{Scott}, A.~D. 2000, \mnras, 313, 775

\bibitem[{{Sekiguchi} {et~al.}(1988){Sekiguchi}, {Feast}, {Whitelock},
  {Overbeek}, {Wargau}, \& {Jones}}]{Sekiguchi1988}
{Sekiguchi}, K., {Feast}, M.~W., {Whitelock}, P.~A., {Overbeek}, M.~D.,
  {Wargau}, W., \& {Jones}, J.~S. 1988, \mnras, 234, 281

\end{thebibliography}

\end{document}